\documentclass[9pt,twoside]{article}
\begin{document}

{\noindent\bf \Large Achievable efficiencies for probabilistically  cloning the states } \\[0.2cm]

\leftskip 2cm

{\noindent\bf T Gao$^{1,2}$,  F L Yan $^{3,4}$ and Z X Wang $^1$}\\[0.2cm]

 {\noindent\footnotesize $^1$ Department of Mathematics, Capital Normal University, Beijing 100037,
 China\\
 $^2$ College of Mathematics and Information Science, Hebei Normal
University, Shijiazhuang
050016, China\\
$^3$ Department of Physics, Hebei Normal University, Shijiazhuang 050016, China\\
$^4$ CCAST (World Laboratory), P.O. Box 8730, Beijing 100080, China\\[0.2cm]
E-mail: gaoting@heinfo.net\\[0.2cm]}
Received ~~~~ November 2003\\[0.2cm]
{{\bf Abstract}\\
\noindent We present an example of quantum computational tasks whose performance is enhanced if we distribute
quantum information using quantum cloning. Furthermore we give  achievable efficiencies for probabilistic
cloning the quantum states used in implemented tasks for which cloning provides some enhancement in
performance.\\[0.5cm]}

\leftskip 0cm

{\noindent{\bf 1. Introduction  }}\\[0.2cm]
Cloning is  a type of quantum information processing  tool. In 1982  Wootters and Zurek \cite {s1} and Dieks
\cite {s2} independently  discovered  the no-cloning theorem, one of the first results stressing the
peculiarities of quantum information. They showed  that unlike classical information,  it is impossible to make
perfect copies of an unknown quantum state, i.e. qubits can not be copied. Since then quantum cloning has been
studied intensively, and
 much effort has been put into developing optimal cloning processes
 [3-14]. There are two main approaches to
quantum cloning. The first one consists of using ancillary quantum systems and a global unitary operation to
obtain multiple imperfect clones of a given, unknown quantum state. These universal quantum cloning machines
(UQCM's) were first invented by Bu$\check{\rm z}$ek and Hillery \cite {s3} and developed by other authors
[4-12]. The  second kind of cloning procedure first designed by  Duan and Guo \cite {s13, s14} is
nondeterministic, consisting in adding an ancilla, performing unitary operations and measurements, with a
postselection of the measurement results. The resulting clones are perfect, but the procedure only succeeds with
a certain probability $p<1$, which depends on the particular set of the states that we are trying to clone.
Recently, Galv$\tilde{\rm a}$o and Hardy discuss how quantum information distribution implemented with different
types of quantum cloning procedures can improve the performance of  some quantum computation tasks \cite {s15}.
Unfortunately in the second example they obtained the achievable efficiencies for probabilistically cloning the
states by  a numerical search. Evidently the numerical result is not an exact solution and this is what
originally motivated the present work.

Our purpose in this paper is twofold. First we present an example of quantum computation tasks whose performance
is enhanced if we distribute quantum information using quantum cloning. The second purpose of the paper is to
provide  achievable efficiencies for probabilistically cloning the states \cite {s15} used in implemented tasks
for which cloning provides some enhancement in performance.\\[0.2cm]
{\noindent\bf 2. An example with probabilistic cloning}\\[0.2cm]
 In this section we give an example of quantum computation tasks that can be better performed if we  make use of quantum
 cloning. The task relies on state-dependent probabilistic quantum cloning discussed by Duan and Guo
 \cite{s13, s14}.
 Now we present our example by generalizing  the second example of Ref. \cite {s15} in which they
 discussed
 the functions that take two bits to one bit,   to the case  of three bits to one bit.

 The quantum computational task  is  as follows. Suppose that we are given
 $3$ quantum black-boxes.  What each black-box does is to accept four $2$-level quantum systems as an input and
 apply a unitary operator to it, producing the evolved state as the output. We take the black-boxes to consist
 of arbitrary quantum circuit that query a given function only once. The query of function $f_i$ is the unitary
 that performs $|x\rangle|y\rangle\rightarrow|x\rangle|y\oplus f_i(x)\rangle$, where the symbol $\oplus$
 represent the bitwise $XOR$ operation. Our task will involve determining two functionals, one depending only on
 $f_0$ and $f_1$, and the other on $f_0$ and $f_2$. We will prove that cloning offers an advantage which cannot
 be matched by any approach that does not resort to quantum cloning.

In order to precisely state our task, we start by considering all functions $h_i$ which take three bits to one
bit. We may represent each such function with eight bits $a_1$, $a_2$, $a_3$, $a_4$, $a_5$, $a_6$, $a_7$, and
$a_8$, writing $h_{a_1a_2a_3a_4a_5a_6a_7a_8}$ to stand for  the function $h$ such that $h(000)=a_1$,
$h(001)=a_2$, $h(010)=a_3$, $h(011)=a_4$, $h(100)=a_5$, $h(101)=a_6$, $h(110)=a_7$, $h(111)=a_8$. Now we define
some sets of functions that will be useful in stating our task:
 \[S_{f_0}=\{h_{01000000}, h_{00110011},
h_{11000011}\},\]
\begin{eqnarray*} &&S_1=\{h_{01000000}, h_{10110000}, h_{10001100}, h_{00100110},
h_{00010101}, h_{10000011}, h_{00101001}, h_{00011010}\},\nonumber\\ &&S_2=\{h_{00000000}, h_{00001111},
h_{01010101}, h_{00110011}, h_{10011001}, h_{11000011}, h_{01101001}, h_{10100101}\},\end{eqnarray*}
\[S_{f_{12}}=S_1\cup S_2,\]
\begin{eqnarray*}
&&S_{00000000}=\{h_{00000000}, h_{11111111}\}, S_{00001111}=\{h_{00001111}, h_{11110000}\},
S_{01010101}=\{h_{01010101}, h_{10101010}\},\\
&&S_{00110011}=\{h_{00110011}, h_{11001100}\}, S_{10011001}=\{h_{10011001}, h_{01100110}\},
S_{11000011}=\{h_{11000011}, h_{00111100}\},\\
&&S_{01101001}=\{h_{01101001}, h_{10010110}\}, S_{10100101}=\{h_{10100101}, h_{01011010}\},
\end{eqnarray*}
\begin{eqnarray*}S_f=S_{00000000}\cup S_{00001111}\cup S_{01010101}\cup
S_{00110011} \cup S_{10011001}\cup S_{11000011}\cup S_{01101001}\cup S_{10100101}.\end{eqnarray*}

Now we first randomly choose  a function $f_0\in S_{f_0}$, then two other functions $f_1$ and $f_2$ are picked
from the set $S_{f_{12}}$, also at  random  but satisfying:
\begin{equation}\label{f123}
f_0\oplus f_1, ~~f_0\oplus f_2\in S_f.
\end{equation}
Here the  symbol $\oplus$  is addition modulo 2. The task will be to find in which of the eight sets
$S_{00000000}$, $S_{00001111}$, $S_{01010101}$, $S_{00110011}$, $S_{10011001}$, $S_{11000011}$, $ S_{01101001}$
and $S_{10100101}$ lie each of the functions $f_0\oplus f_1$ and $f_0\oplus f_2$, applying  quantum circuits
that query $f_0$, $f_1$, and $f_2$ at most once each.  Our score will be given by the average probability of
successfully guessing both correctly.\\[0.2cm]
{\it 2.1 Score without cloning}\\[0.2cm]
Now we will give the attainable score if we do not resort to  cloning.  Just as \cite{s15} the best no-cloning
strategy goes as follows. Firstly, from the constraints given by Eq.(\ref{f123}) we  note that  both $f_1$ and
$f_2$ must be in $S_1$ if  $f_0=h_{01000000}$, and  $f_1$ and $f_2$ must belong to $S_2$ if  $f_0$ is either
$h_{00110011}$ or $h_{11000011}$.
 Since $f_0$ were drawn from a uniformly random
distribution, the probability of  both $f_1$ and $f_2$  in $S_2$
is $2/3$. Assume that it is the case, then we can discriminate
between the two possibilities for $f_0$ with a single, classical
function call. Furthermore, by using the quantum circuit in Fig.1
twice (once each with $f_1$ and $f_2$) we can distinguish the
eight possibilities for functions $f_1$ and $f_2$.

This happens because depending on which function in $S_2$ was queried, this quantum circuit results in one of
the eight orthogonal states
\begin{equation}
|\varphi_i\rangle=\frac {1}{2\sqrt 2}\sum_{x=000}^{111}(-1)^{f_i(x)}|x\rangle.
 \end{equation}
This allows us to determine functions $f_0$, $f_1$, and $f_2$ correctly with probability $p=2/3$, in which case
we can determine which sets contain $f_0+f_1$ and $f_0+f_2$ and accomplish our task. Even in the case where the
initial assumption about $f_0$ was wrong, we may still have guessed the right sets by chance; the chances of
getting both right this way are $1/64$. Thus, the best no-cloning average score is
\begin{equation}
p_1=\frac {2}{3} + \frac {1}{3}\cdot\frac {1}{64} =0.671875.
\end{equation}\\[0.2cm]
{\it 2.2 Score with cloning }\\[0.2cm]
  Next we will prove that  we can do much better than that with quantum cloning.
  The idea is similar to  Ref. \cite {s15},
   that  is,
  to devise a quantum circuit that queries function $f_0$ only
  once, makes two clones of the resulting state, and then queries
  functions $f_1$ and $f_2$, one in each branch of the
  computation. Since we have some information about the state
  produced by one query of $f_0$, the probabilistic cloning machines investigated
   by Duan and Guo \cite {s13} will suit this task better.

The quantum circuit that we use to solve this problem is depicted
in Fig.2.

 Immediately after querying function $f_0$, we have one of three
possible linearly independent states (each corresponding to one of the possible $f_0$'s):
{\small\begin{eqnarray} && |\Psi_1\rangle\equiv|h_{01000000}\rangle \equiv \frac {1}{2\sqrt
{2}}[|000\rangle-|001\rangle+|010\rangle+ |011\rangle+|100\rangle+|101\rangle
+|110\rangle+|111\rangle],\label{h11}\\
&& |\Psi_2\rangle\equiv|h_{00110011}\rangle \equiv \frac {1}{2\sqrt {2}}[|000\rangle+|001\rangle-|010\rangle-
|011\rangle+|100\rangle+|101\rangle-|110\rangle-|111\rangle],\label{h12}\\
 && |\Psi_3\rangle\equiv|h_{11000011}\rangle \equiv \frac {1}{2\sqrt {2}}[-|000\rangle-|001\rangle+|010\rangle+
|011\rangle+|100\rangle +|101\rangle-|110\rangle-|111\rangle].\label{h13}
\end{eqnarray}}

 The probabilistic cloning machines with different cloning efficiencies (defined as the probability of cloning
successfully) for each of states \ref{h11}--\ref{h13} will be constructed.  From Theorem 2 in Ref. \cite{s13} we
obtain the following exact achievable efficiencies
\begin{eqnarray}
 &\gamma_1\equiv \gamma(|h_{01000000}\rangle)=\frac {7}{127},\label{r11}\\
 &\gamma_2\equiv \gamma(|h_{00110011}\rangle)=\gamma_3\equiv \gamma(|h_{11000011})=\frac {112}{127},\label{r123}
\end{eqnarray}
which will be shown in next section.

After the cloning process a measurement on a "flag" subsystem is
performed and the result will tell us  whether the cloning was
successful or not. For this particular cloning process, the
probability of success is, on average, $P_{\rm
success}=(\gamma_1+\gamma_2+\gamma_3)/3=\frac {77}{127}$. If it
was successful, then each of the cloning branches goes through the
second part of the circuit in Fig.2 , to yield one of the eight
orthogonal states:
\begin{eqnarray}
&& |h_{00000000}\rangle\equiv \frac {1}{2\sqrt 2}[|000\rangle+|001\rangle+|010\rangle+|011\rangle+|100\rangle
+|101\rangle+|110\rangle+|111\rangle],\label{h1}\\
&& |h_{00001111}\rangle\equiv \frac {1}{2\sqrt 2}[|000\rangle+|001\rangle+|010\rangle+|011\rangle-|100\rangle
-|101\rangle-|110\rangle-|111\rangle],\label{h2}\\
&& |h_{01010101}\rangle\equiv \frac {1}{2\sqrt
2}[|000\rangle-|001\rangle+|010\rangle-|011\rangle+|100\rangle-|101\rangle+|110\rangle-|111\rangle],\\
&& |h_{00110011}\rangle\equiv \frac {1}{2\sqrt
2}[|000\rangle+|001\rangle-|010\rangle-|011\rangle+|100\rangle+|101\rangle-|110\rangle-|111\rangle],\\
&& |h_{10011001}\rangle\equiv \frac {1}{2\sqrt
2}[-|000\rangle+|001\rangle+|010\rangle-|011\rangle -|100\rangle+|101\rangle+|110\rangle-|111\rangle],\\
&& |h_{11000011}\rangle\equiv \frac {1}{2\sqrt
2}[-|000\rangle-|001\rangle+|010\rangle+|011\rangle +|100\rangle+|101\rangle-|110\rangle-|111\rangle],\\
&& |h_{01101001}\rangle\equiv \frac {1}{2\sqrt 2}[|000\rangle-|001\rangle-|010\rangle+|011\rangle-|100\rangle
+|101\rangle+|110\rangle-|111\rangle],\\
&& |h_{10100101}\rangle\equiv \frac {1}{2\sqrt 2}[-|000\rangle+|001\rangle-|010\rangle+|011\rangle
+|100\rangle-|101\rangle+|110\rangle-|111\rangle],\label{h8}
\end{eqnarray}
which can be discriminated unambiguously.  Therefore, if the
cloning process is successful, we manage to accomplish our task.

However, the cloning process may fail with  probability $(1-P_{\rm success})$. If this happens, it is more
likely to be $h_{01000000}$ than the other two, because of  the relatively low cloning efficiency for the state
in Eq.(\ref{h11}), in relation to the states in Eqs.(\ref{h12}) and (\ref{h13}) [see Eqs. (\ref{r11}) and
(\ref{r123})]. If we then guess that $f_0=h_{01000000}$, we will be right with probability
\begin{equation}
p_{01000000}=\frac {(1-\gamma_1)}{(1-\gamma_1)+(1-\gamma_2)+(1-\gamma_3)}=\frac {4}{5}.
\end{equation}
What is more, we are still free to design quantum circuits to
obtain information about $f_1$ and $f_2$, since at this stage we
still have not queried them. Given our guess that
$f_0=h_{01000000}$, only the eight functions in $S_1$ can be
candidates for $f_1$ and $f_2$, because of the constraints given
by Eq.(\ref{f123}). These eight possibilities can be
discriminated unambiguously by run a circuit like that of Fig.1
twice, once with $f_1$ and once with $f_2$. The circuit produces
one of eight orthogonal states, each corresponding to one of the
eight possibilities for $f_i$. Therefore if our  guess that
$f_0=h_{01000000}$ was correct, we are able to find the correct
$f_1$ and $f_2$ and therefore accomplish our task. In the case
that $f_0\neq h_{01000000}$ after all, we may still have guessed
the right sets by chance; a simple analysis shows that this will
happen with probability 1/64.

The above considerations leads to an overall probability of
success given by
\begin{eqnarray}\label{y}
p_2&&=P_{\rm success}+(1-P_{\rm success})[p_{01000000}+(1-p_{01000000})\frac
{1}{64}]\nonumber\\
&&=\frac
{22+21(\gamma_2+\gamma_3)}{64}\nonumber\\
&&=\frac {3749}{4064}\nonumber\\
&&\simeq 0.92249\nonumber\\
&&>p_1=0.671875,
\end{eqnarray}
 thus showing
that this cloning approach is more efficient than the previous one, which does not use cloning.\\[0.2cm]
{\it 2.3 Exact achievable efficiencies }\\[0.2cm]
 Here we present the analytic solution of achievable efficiencies for cloning the state Eqs. (\ref{h11})-(\ref{h13}).
 As stated above we use  $\gamma_1\equiv
\gamma(|h_{01000000}\rangle)$, $\gamma_2\equiv \gamma(|h_{00110011}\rangle)$,  $\gamma_3\equiv
\gamma(|h_{11000011}\rangle)$ to express the achievable efficiencies, and let
 $|P^{(1)}\rangle$, $|P^{(2)}\rangle$, $|P^{(3)}\rangle$  be  normalized states of the flag $P$. $P_{ij}$
  denotes the inner product $\langle P^{(i)}|P^{(j)}\rangle$ between $|P_i\rangle$ and $|P_j\rangle$, $i, j=1, 2, 3$.
   Clearly, $|P_{ij}|\leq 1$.
    Suppose the $3\times 3$ matrices
$X^{(1)}=[\langle\Psi_i|\Psi_j\rangle]$,  $X^{(2)}_P=[\langle\Psi_i|\Psi_j\rangle^2P_{ij}]$ and the diagonal
efficiency matrix  $\Gamma={\rm diag}(\gamma_1,\gamma_2,\gamma_3)$,
 then
{\small\begin{eqnarray*} X^{(1)}-\sqrt{\Gamma}X^{(2)}_P\sqrt{\Gamma^+}
 &&=\left(
\begin{array}{ccc}
1&-\frac{1}{4}&\frac{1}{4}\\
-\frac{1}{4}&1&0\\
\frac{1}{4}&0&1 \end{array}\right)- \left(
\begin{array}{ccc}
\gamma_1&\frac{\sqrt{\gamma_1\gamma_2}}{16}P_{12}&\frac{\sqrt{\gamma_1\gamma_3}}{16}P_{13}\\
\frac{\sqrt{\gamma_1\gamma_2}}{16}P_{12}^*&\gamma_2&0\\
\frac{\sqrt{\gamma_1\gamma_3}}{16}P_{13}^*&0&\gamma_3
\end{array}\right)\\
&&=\left(
\begin{array}{ccc}
1-\gamma_1&-\frac{1}{4}-\frac{\sqrt{\gamma_1\gamma_2}}{16}P_{12}&\frac{1}{4}-\frac{\sqrt{\gamma_1\gamma_3}}{16}P_{13}\\
-\frac{1}{4}-\frac{\sqrt{\gamma_1\gamma_2}}{16}P_{12}^*&1-\gamma_2&0\\
\frac{1}{4}-\frac{\sqrt{\gamma_1\gamma_3}}{16}P_{13}^*&0&1-\gamma_3
\end{array}\right)
\end{eqnarray*}}

Theorem 2 of Ref.\cite {s13} provides us with inequalities {\small\begin{eqnarray}
&&1-\gamma_1\geq 0,~~~~~~~~~~~~~~~~~~~~\label{p11}\\
&&(1-\gamma_1)(1-\gamma_2)-|\frac {1}{4}+\frac {1}{16}\sqrt {\gamma_1\gamma_2}P_{12}|^2\geq
 0,~~~~~~~~~~~~~~~~~~~\label{p12}\\
&&(1-\gamma_1)(1-\gamma_2)(1-\gamma_3)-(1-\gamma_3)|\frac {1}{4}+\frac {1}{16}\sqrt
{\gamma_1\gamma_2}P_{12}|^2-(1-\gamma_2)|\frac {1}{4} -\frac {1}{16}\sqrt {\gamma_1\gamma_3}P_{13}|^2\geq
0,\label{p13}
\end{eqnarray}}which allow us to derive achievable efficiencies for the probabilistic cloning process. According to the rule
stated in above section (see Eq. (\ref {y}))  the overall probability (score) of success with the help of
probabilistic cloning is given by {\small\begin{eqnarray}
p_2&=&p_{\rm success}+(1-p_{\rm success})[p_{01000000}+(1-p_{01000000})\frac{1}{64}] \nonumber\\
   &=&\frac{\gamma_1+\gamma_2+\gamma_3}{3}+(1-\frac{\gamma_1+\gamma_2+\gamma_3}{3})
   [\frac{1-\gamma_1}{3-\gamma_1-\gamma_2-\gamma_3}+(1-\frac{1-\gamma_1}{3-\gamma_1-\gamma_2-\gamma_3})\frac{1}{64}]\nonumber\\
   &=&[22+21(\gamma_2+\gamma_3)]/64.
\end{eqnarray}}From above equation we know that we should find the maximum of $\gamma_2+\gamma_3$ satisfying
Eqs.(\ref{p11})--(\ref{p13}).

In the following, we show that the maximum of $\gamma_2+\gamma_3$ must be greater than or equal to
$\frac{224}{127}$. We consider the case $\gamma_2=\gamma_3$. In this case, there is {\small
\begin{equation} (1-\gamma_1)(1-\gamma_2)-|\frac {1}{4}+\frac {1}{16}\sqrt {\gamma_1\gamma_2}P_{12}|^2-|\frac
{1}{4}-\frac {1}{16}\sqrt {\gamma_1\gamma_2}P_{13}|^2\geq 0,
\end{equation}}which implies that
\begin{equation}\label{xy11}
 \frac{7}{8}-qx+sx^2\geq y\geq 2x\geq 0,
\end{equation}
where $P_{12}=a+b{\rm i}$, $P_{13}=c+d{\rm i}$, $q=\frac{1}{32}(a-c)$, $s=1-\frac{1}{256}(a^2+b^2+c^2+d^2)$,
$y=\gamma_1+\gamma_2$, and $x=\sqrt{\gamma_1\gamma_2}$. It is not difficult to prove that
\begin{equation}\label{sq}
    \frac{127}{128}\leq s\leq 1, ~~~ -\frac{1}{16}\leq q\leq \frac{1}{16}.
\end{equation}
  Since
$\frac{7}{8}-q+s\leq 2$, and $0\leq x\leq 1$,
$y=\frac{7}{8}-qx+sx^2$ and $y=2x$ have one intersection point
{\small $${ (x_0,y_0)=\biggl(\frac {2+q-\sqrt
{(2+q)^2-\frac{7}{2}s}}{2s},\frac {2+q-\sqrt
{(2+q)^2-\frac{7}{2}s}}{s}\biggr).}$$}The region in $x$-$y$ plane
and the region in $q$-$s$ plane governed by Eq.(\ref{xy11}) are
the shaded area  in Fig.3 and in Fig.4 respectively.

From $y=\gamma_1+\gamma_2$ and $x=\sqrt{\gamma_1\gamma_2}$ we have
\begin{equation}
\gamma_1=\frac {1}{2}(y-\sqrt {y^2-4x^2}),~~~  \gamma_2=\frac {1}{2}(y+\sqrt {y^2-4x^2}).
\end{equation}
This implies that $\gamma_2$ is a decreasing function of $x$ when $y$ is definite, so the maximum of $\gamma_2$
should occur in the curve
\begin{equation}
\frac {7}{8}-qx+sx^2=y,
\end{equation}
that is, the maximum of $\gamma_2$ must be the point such that $\frac{d\gamma_2}{dx}=\frac{\partial
\gamma_2}{\partial y}\frac{dy}{dx}+\frac{\partial\gamma_2}{\partial x}\frac{dx}{dx}=0$ ( i.e. $x_1=\frac
{\frac{7}{2}s+q^2-4+\sqrt {(\frac{7}{2}s+q^2-4)^2-14sq^2}}{4sq}$, $y_1=\frac {7}{8}-qx_1+sx_1^2$.) Thus,
 the maximum of $\gamma_2$ in the plane
$\gamma_2=\gamma_3$ is
\begin{equation}
\gamma_2=\frac {1}{2}\biggl\{\frac {7}{8}-qx_1+sx_1^2+\sqrt{(\frac {7}{8}-qx_1+sx_1^2)^2-4x_1^2}\biggr\},
\end{equation}
where
\begin{equation}
x_1=\frac {\frac{7}{2}s+q^2-4+\sqrt {(\frac{7}{2}s+q^2-4)^2-14sq^2}}{4sq}.
\end{equation}
Let
\begin{equation}\label{} w=w(q,s)=\frac {7}{8}-qx_1+sx_1^2; ~~~ v=v(q,s)=x_1,
\end{equation}
then  $w_{s=1-2q^2}=\frac{9}{16}\sqrt{49+32v^2}-\frac{49}{16}$
when $s=1-2q^2$;
$v_{s=\frac{127}{128}}=\frac{q^2-\frac{135}{256}+\sqrt{q^4-\frac{1913}{128}q^2+(\frac{135}{256})^2}}{\frac{127}{32}q}$
and
$w_{s=\frac{127}{128}}=\frac{7}{8}-\frac{q^2-\frac{135}{256}+\sqrt{q^4-\frac{1913}{128}q^2+(\frac{135}{256})^2}}
{\frac{127}{32}}+\frac{8[q^2-\frac{135}{256}+\sqrt{q^4-\frac{1913}{128}q^2+(\frac{135}{256})^2}]^2}{127q^2}$
when $s=\frac{127}{128}$. The $(v,w)$ region corresponding $(q,s)$
region in Fig.4 is depicted in Fig.5.

Because $\gamma_2$ is a decreasing function of $v$ while $w$ is definite, the maximum of $\gamma_2$ must be in
the left boundary curve $w_{s=\frac{127}{128}}$ in $v$-$w$ plane corresponding to the boundary
$s=\frac{127}{128}$ in $q$-$s$ plane. By $\frac {{\rm d}\gamma_2}{{\rm d}q}<0$, the maximum of $\gamma_2$ should
be at the point
\begin{equation}\label{}
    q=-\frac{1}{16},~~~ s=\frac{127}{128}.
\end{equation}

The exact maximum of $\gamma_2$ is
\begin{eqnarray}
&&\gamma_2\equiv \gamma(|h_{00110011}\rangle)=\gamma(|h_{11000011}\rangle)=\frac {112}{127},\\
&&\gamma_1\equiv \gamma(|h_{01000000}\rangle)=\frac {7}{127}.
\end{eqnarray}

 So we do find an exact solution of achievable efficiencies $\gamma_1, \gamma_2, \gamma_3$ satisfying
 $\gamma_2=\gamma_3$, and prove that the maximum $\gamma_2+\gamma_3$ must be greater than or equal to $\frac
 {224}{127}$.\\[0.3cm]
{\bf 3. Exact achievable efficiencies for probabilistically cloning the states of Ref. \cite {s15}}\\[0.3cm]
In this section we will give the exact achievable efficiencies for probabilistically cloning the states in the
second example of Ref. \cite {s15}.

 In Ref. \cite {s15}, the probabilistic cloning quantum states are
\begin{eqnarray}
&&|h_1\rangle=|h_{0010}\rangle\equiv\frac
{1}{2}[|00\rangle+|01\rangle-|10\rangle+|11\rangle],\label{h21}\\
&&|h_2\rangle=|h_{0101}\rangle\equiv\frac
{1}{2}[|00\rangle-|01\rangle+|10\rangle-|11\rangle],\label{h22}\\
&&|h_3\rangle=|h_{1001}\rangle\equiv\frac {1}{2}[-|00\rangle+|01\rangle+|10\rangle-|11\rangle].\label{h23}
\end{eqnarray}
We can build probabilistic cloning machines with different cloning efficiencies  for each of the states
\ref{h21}--\ref{h23}.      Let $\gamma_1\equiv \gamma(|h_{0010}\rangle)$, $\gamma_2\equiv
\gamma(|h_{0101}\rangle)$,  $\gamma_3\equiv \gamma(|h_{1001}\rangle)$ be the achievable efficiencies, and
 $|P^{(1)}\rangle$, $|P^{(2)}\rangle$, $|P^{(3)}\rangle$  be  normalized states of the flag $P$. $P_{ij}$
  denotes the inner product between $|P_i\rangle$ and $|P_j\rangle$, $i, j=1, 2, 3$. Clearly, $|P_{ij}|\leq 1$.
    Suppose
\begin{eqnarray*}
&&X^{(1)}=\left(
\begin{array}{ccc}
\langle h_1|h_1\rangle&\langle h_1|h_2\rangle&\langle
h_1|h_3\rangle\\
\langle h_2|h_1\rangle&\langle h_2|h_2\rangle&\langle
h_2|h_3\rangle\\
\langle h_3|h_1\rangle&\langle h_3|h_2\rangle&\langle h_3|h_3\rangle
\end{array}\right),\\
&&X^{(2)}_P=\left(
\begin{array}{ccc}
\langle h_1|h_1\rangle^2 P_{11}&\langle h_1|h_2\rangle^2 P_{12}&\langle
h_1|h_3\rangle^2 P_{13}\\
\langle h_2|h_1\rangle^2 P_{21}&\langle h_2|h_2\rangle^2 P_{22}&\langle
h_2|h_3\rangle^2 P_{23}\\
\langle h_3|h_1\rangle^2 P_{31}&\langle h_3|h_2\rangle^2 P_{32}&\langle h_3|h_3\rangle^2 P_{33}
\end{array}\right),\\
&&\sqrt{\Gamma}=\left(
\begin{array}{ccc}
\sqrt{\gamma_1}&0&0\\
0&\sqrt{\gamma_2}&0\\
0&0&\sqrt{\gamma_3} \end{array}\right) \end{eqnarray*} then
{\small\begin{eqnarray*}
&&~~~~~X^{(1)}-\sqrt{\Gamma}X^{(2)}_P\sqrt{\Gamma^+} \\
&&=\left(
\begin{array}{ccc}
1-\gamma_1&-\frac{1}{2}-\frac{\sqrt{\gamma_1\gamma_2}}{4}P_{12}&-\frac{1}{2}-\frac{\sqrt{\gamma_1\gamma_3}}{4}P_{13}\\
-\frac{1}{2}-\frac{\sqrt{\gamma_1\gamma_2}}{4}P_{12}^*&1-\gamma_2&0\\
-\frac{1}{2}-\frac{\sqrt{\gamma_1\gamma_3}}{4}P_{13}^*&0&1-\gamma_3
\end{array}\right)
\end{eqnarray*}}

Theorem 2 of Ref.\cite {s13} provides us with inequalities
{\small\begin{eqnarray}
&&1-\gamma_1\geq 0,~~~~~~~~~~~~~~~~~~~~\label{p21}\\
&&(1-\gamma_1)(1-\gamma_2)-|\frac {1}{2}+\frac {1}{4}\sqrt {\gamma_1\gamma_2}P_{12}|^2\geq
0,~~~~~~~~~~~~~~~~~~~\label{p22}\\
 &&(1-\gamma_1)(1-\gamma_2)(1-\gamma_3)-(1-\gamma_3)|\frac {1}{2}+\frac
{1}{4}\sqrt {\gamma_1\gamma_2}P_{12}|^2 -(1-\gamma_2)|\frac {1}{2}+\frac {1}{4}\sqrt
{\gamma_1\gamma_3}P_{13}|^2\geq 0,\label{p23}
\end{eqnarray}}which allow us to derive achievable efficiencies for the probabilistic cloning process.   According to the rule
specified in Ref.\cite {s15} the overall probability (score) of success with the help of probabilistic cloning
is given by
\begin{eqnarray}
p_2&=&p_{\rm success}+(1-p_{\rm success})[p_{0010}+(1-p_{0010})\frac{1}{16}] \nonumber\\
   &=&\frac{\gamma_1+\gamma_2+\gamma_3}{3}+(1-\frac{\gamma_1+\gamma_2+\gamma_3}{3})[\frac{1-\gamma_1}{3-\gamma_1-\gamma_2-\gamma_3}+(1-\frac{1-\gamma_1}{3-\gamma_1-\gamma_2-\gamma_3})\frac{1}{16}]\nonumber\\
   &=&[6+5(\gamma_2+\gamma_3)]/16.
\end{eqnarray}
From above equation we know that we should find the maximum of $\gamma_2+\gamma_3$ satisfying Eqs.
(\ref{p21})-(\ref{p23}).

Our immediate goal is to prove that the maximum of $\gamma_2+\gamma_3$ must be greater than or equal to 8/7. For
this purpose we discuss the problem in the plane $\gamma_2=\gamma_3$. In this plane Eq. (\ref {p23}) becomes
\begin{equation}
(1-\gamma_1)(1-\gamma_2)-|\frac {1}{2}+\frac {1}{4}\sqrt {\gamma_1\gamma_2}P_{12}|^2-|\frac {1}{2}+\frac
{1}{4}\sqrt {\gamma_1\gamma_2}P_{13}|^2\geq 0.\label{sp23}
\end{equation}
Let
\begin{eqnarray}
&&P_{12}=a+b{\rm i}, ~~~ P_{13}=c+d{\rm i},
 ~~~q=\frac {1}{4}(a+c), s=1-\frac
{1}{16}(a^2+b^2+c^2+d^2), x=\sqrt {\gamma_1\gamma_2},~~~ y=\gamma_1+\gamma_2.
\end{eqnarray}
 Then Eq. (\ref{sp23}) can be rewritten concisely as
\begin{equation}
\frac {1}{2}-qx+sx^2\geq y.
\end{equation}
Obviously
\begin{equation}
\frac {1}{2}-qx+sx^2\geq y\geq 2x\geq 0.\label{xy22}
\end{equation}
Here $y=\frac {1}{2}-qx+sx^2$ and $y=2x$ have one intersection point
\begin{eqnarray}
x_0=\frac {2+q-\sqrt {(2+q)^2-2s}}{2s},~~~~ y_0=2x_0.
\end{eqnarray}
 The proof is as follows: The intersection points of
$y=\frac {1}{2}-qx+sx^2=\frac{1}{2}-\frac{1}{4}(c+a)x+[1-\frac{1}{16}(a^2+b^2+c^2+d^2)]x^2$ and $y=2x$ are
$x_0=\frac {2+q\pm\sqrt {(2+q)^2-2s}}{2s}, y_0=2x_0$. From $|P_{12}|\leq 1$ and $|P_{13}|\leq 1$ it is seen
$|a+c|\leq 2$ and $0\leq a^2+b^2+c^2+d^2\leq 2$, which imply that
\begin{equation}
-\frac{1}{2}\leq q\leq \frac{1}{2}, ~~~\frac{7}{8}\leq s\leq 1,
\end{equation}
thus $x_0=\frac {2+q+\sqrt {(2+q)^2-2s}}{2s}> 1$ that contradict with $x=\sqrt{\gamma_1\gamma_2}\leq 1$.
Therefore $y=\frac {1}{2}-qx+sx^2$ and $y=2x$ have one intersection point $x_0=\frac {2+q-\sqrt
{(2+q)^2-2s}}{2s}, y=2x_0$.

The region in $x-y$ plane governed by Eq. (\ref{xy22})
 is shown in Fig.6,
 where  $x$ must satisfy
\begin{equation}
0\leq x\leq\frac {2+q-\sqrt {(2+q)^2-2s}}{2s}=x_0.
\end{equation}

Immediately $\frac{\partial x_0}{\partial
q}=\frac{1}{2s}[1-\frac{2+q}{\sqrt{(2+q)^2-2s}}]\leq 0$. It
follows that when $s$ is definite $x_0$ is a decreasing function
of $q$. If $q$ is definite (i.e. $a+c=k$ is definite), then the
maximum $s$ is to make $a^2+b^2+c^2+d^2=(a+c)^2+b^2+d^2-2ac$
minimum, which imply $b=d=0$ and $ac=\frac{(a+c)^2}{4}$. Therefore
the curve of maximum $s$ is $s=1-\frac{1}{2}q^2$ when $q$ is
definite. While $s$ minimum is to make $a^2+b^2+c^2+d^2$ maximum,
so minimum $s$ is  $s=\frac{7}{8}$ in the case $q$ is definite.
The boundary of $s$ and $q$ is illustrated in Fig.7.

 By $x=\sqrt {\gamma_1\gamma_2}$ and
$y=\gamma_1+\gamma_2$ we get
\begin{equation}
\gamma_1=\frac {1}{2}(y-\sqrt {y^2-4x^2}),~~~  \gamma_2=\frac {1}{2}(y+\sqrt {y^2-4x^2}).
\end{equation}
It follows that if $y$ is definite, the smaller $x$ is, the bigger $\gamma_2$ is, so the maximum of $\gamma_2$
should take place in the curve
\begin{equation}
\frac {1}{2}-qx+sx^2=y,
\end{equation}
that is, the maximum of $\gamma_2$ must be the point such that $\frac{d\gamma_2}{dx}=\frac{\partial
\gamma_2}{\partial y}\frac{dy}{dx}+\frac{\partial\gamma_2}{\partial x}\frac{dx}{dx}=0$ (i.e. $x_1=\frac
{2s+q^2-4+\sqrt {(2s+q^2-4)^2-8sq^2}}{4sq}$, $y_1=\frac {1}{2}-qx_1+sx_1^2$.)  Thus,
 the maximum of $\gamma_2$ in the plane
$\gamma_2=\gamma_3$ is
\begin{equation}
\gamma_2=\frac {1}{2}\{\frac {1}{2}-qx_1+sx_1^2+\sqrt{(\frac {1}{2}-qx_1+sx_1^2)^2-4x_1^2}\},\label{r2x1}
\end{equation}
where
\begin{equation}
x_1=\frac {2s+q^2-4+\sqrt {(2s+q^2-4)^2-8sq^2}}{4sq}.
\end{equation}

Next  we derive  the maximum of $\gamma_2$. Let
\begin{equation}
 w=w(q,s)=\frac {1}{2}-qx_1+sx_1^2; ~~~~ v=v(q,s)=x_1.
\end{equation}
Now we change $(q,s)$ region to $(v,w)$ region. When
$s=1-\frac{1}{2}q^2$, then $v=-\frac{q}{2-q^2}$ and
$w=\frac{1}{2}+\frac{3q^2}{2(2-q^2)}$. From $v=-\frac{q}{2-q^2}$,
$|q|\leq \frac{1}{2}$ and $v=x_1\geq 0$ we know that
$q=\frac{1-\sqrt{1+8v^2}}{2v}$, $0\leq v\leq \frac{2}{7}$. Hence
$w_{s=1-\frac{q^2}{2}}=-\frac{1}{4}+\frac{3}{4}\sqrt{1+8v^2}$ and
$0\leq v_{s=1-\frac{q^2}{2}}\leq \frac{2}{7}$ in the case
$s=1-\frac{1}{2}q^2$. Note
$v_{s=\frac{7}{8}}=\frac{-\frac{9}{4}+q^2+\sqrt{(q^2-\frac{9}{4})^2-7q^2}}{\frac{7}{2}q}$
and $w_{s=\frac{7}{8}}=\frac
{1}{2}-\frac{-\frac{9}{4}+q^2+\sqrt{(q^2-\frac{9}{4})^2-7q^2}}{\frac{7}{2}}+\frac{7}{8}(\frac{-\frac{9}{4}+q^2+\sqrt{(q^2-\frac{9}{4})^2-7q^2}}{\frac{7}{2}q})^2$
if $s=\frac{7}{8}$. The $(v,w)$ region corresponding $(q,s)$
region is shown in Fig.8.

 Since $\gamma_2$ is a decreasing function of $v$ as $w$ is
 definite, from Eq.(\ref{r2x1}) we obtain that the
maximum of $\gamma_2$ must appear in the left boundary curve $w_{s=\frac{7}{8}}$ in $v-w$ plane corresponding to
the boundary $s=\frac {7}{8}$ in $q-s$ plane. It can be seen that
\begin{equation}
\frac {{\rm d}\gamma_2}{{\rm d}q}<0,
\end{equation}
while $s=\frac {7}{8}$. Therefore the maximum of $\gamma_2$ should exist  at the point
\begin{equation}
q=-\frac {1}{2}, ~~~s=\frac {7}{8}. \end{equation} The exact maximum of $\gamma_2$ is
\begin{equation}
 \gamma_2=\frac {4}{7}\simeq 0.57143\label{r2}
\end{equation}
and
\begin{equation} \gamma_1=\frac {1}{7}\simeq 0.14286.\label{r1}
\end{equation}
It is  clear that our analytic solution is better as compared with the numerical result
\begin{equation}
\gamma_1=0.14165, ~~~~ \gamma_2=\gamma_3=0.57122
\end{equation}
of Ref. \cite {s15}, since Eqs.(\ref{r2}) and (\ref{r1}) are exact solution. Evidently the maximum of $\gamma_2
+ \gamma_3$ should be greater than or equal to $\frac {8}{7}$ although we guess that $\frac {8}{7}$ should be
the maximum of $\gamma_2+\gamma_3$.

However if we make $\gamma_1+\gamma_2$ to be maximum, under the condition $\gamma_2=\gamma_3$, it is not
difficult to obtain that
 the probability of cloning success is, on average,
 \begin{equation}
P_{\rm success}=\gamma_1=\gamma_2=\gamma_3=1-\frac {2\sqrt 2+1}{7}\simeq 0.45308.
\end{equation}

We have constructed the quantum logic network for probabilistically cloning the states \cite {s15} in \cite
{s16}.

 In summary we give  achievable efficiencies for probabilistic cloning
the quantum states used in implemented tasks for which cloning provides some enhancement in performance, and
present an example of quantum computational tasks whose performance is enhanced if we distribute quantum
information using quantum cloning. We hope our result will be helpful in the quantum information
processing.\\[0.2cm]
{\bf Acknowledgments}\\[0.2cm]  This work was supported  by National Natural Science Foundation of China under
Grant No. 10271081 and Hebei Natural Science Foundation under Grant No. 101094.\\

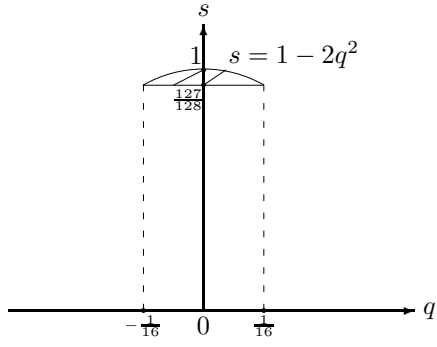
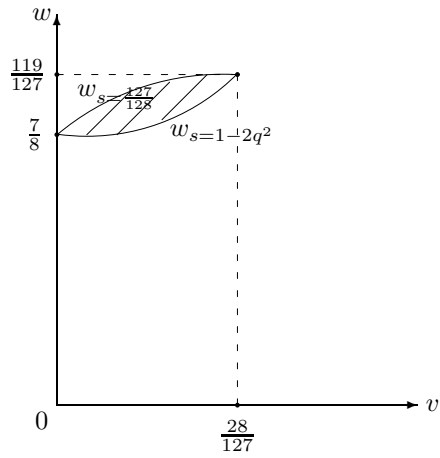
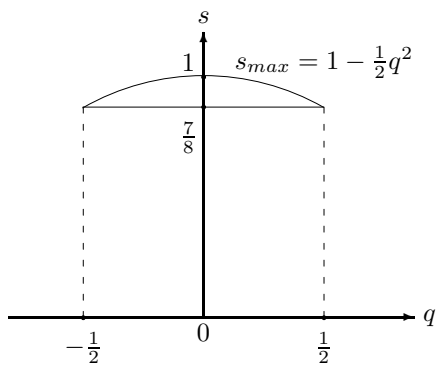
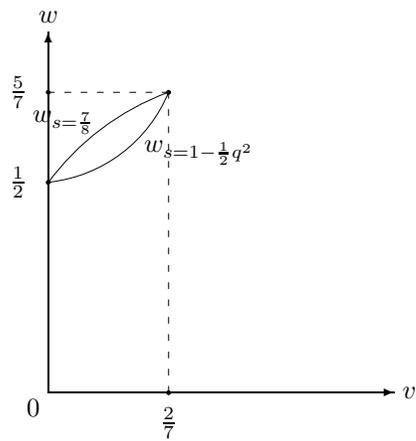
\end{document}